\documentclass[12pt]{iopart}
\usepackage{graphicx}  
\usepackage{bm}        
\usepackage{verbatim}   
\usepackage{iopams}
\usepackage{dcolumn}
\usepackage{epstopdf}
\usepackage{eqnarray}
\usepackage{gensymb}
\usepackage{subcaption}
\usepackage[colorlinks=true, citecolor=red, linkcolor=blue,urlcolor=blue ]{hyperref}

\begin{document}
	
	\title{Active particles in periodic lattices}

\author{Alexander Chamolly$^1$, Takuji Ishikawa$^2$ and Eric Lauga$^1$}

\address{$^1$ Department of Applied Mathematics and Theoretical Physics, University of Cambridge, Wilberforce Road, Cambridge CB3 0WA, United Kingdom}
\address{$^2$  {Department of Finemechanics}, 
	Graduate School of Engineering, 
	Tohoku University, 
	6-6-01, Aoba, Aoba-ku, Sendai 980-8579, Japan}
\eads{\mailto{ajc297@damtp.cam.ac.uk}, \mailto{ishikawa@pfsl.mech.tohoku.ac.jp}, \mailto{e.lauga@damtp.cam.ac.uk}}

\begin{abstract}
	Both natural and artificial small-scale swimmers may often self-propel in environments subject to complex geometrical constraints. While most past theoretical work on low-Reynolds number locomotion addressed idealised geometrical situations, not much  is known on the motion of swimmers in heterogeneous environments. As a first theoretical model, we investigate numerically the behaviour of a single spherical  micro-swimmer located in an infinite, periodic body-centred cubic lattice consisting of rigid inert spheres of the same size as the swimmer. Running a large number of simulations we uncover the phase diagram of possible  trajectories as a function of the strength of the swimming actuation and the packing density of the lattice. We then  use hydrodynamic theory to rationalise our computational results and show in particular how the far-field nature of the swimmer (pusher vs.~puller) governs even the behaviour at high volume fractions. 
\end{abstract}
\pacs{147.15.G-, 47.63.Gd, 47.56.+r}
\vspace{12pt}
	
	\maketitle

\section{Introduction}
Swimming microorganisms live in a  variety of natural and industrial environments, including  the ocean, soil,   intestinal tract and bioreactors, and they play   diverse and important roles in  environmental, agricultural and health issues   \cite{1,2,3}. Understanding the spreading of cells in  various environments is therefore a fundamental problem, and this knowledge can be exploited  for predicting and controlling  the distribution of cells and their growth  \cite{4,5}.

From a physical point of view, the simplest environment for swimming cells is an unbounded fluid in the absence of background flow or the biological responses to stimulus, including chemotaxis, phototaxis or gravitaxis.   The spreading of individual cells in such a simple environment has already been studied thoroughly. Early work   investigated run and tumble motions of individual \emph{Escherichia coli} (\emph{E.~coli}) bacteria and  showed that the  motion of the cells could be  described by random walk models  with the entire population displaying diffusive behaviour  \cite{6}. Similar results were also found for the unicellular alga \emph{Chlamydomonas}, and cell populations displayed long-time  diffusive behaviour \cite{7,8,9,10}.

The behaviour of swimming microorganisms in the presence of a planar wall has also been investigated extensively. Rothschild first reported a significant increase of cell density near a glass wall using bull spermatozoa \cite{11}. Similar accumulation of cells near a wall was also found for human spermatozoa \cite{12} and \emph{E.~coli} bacteria \cite{13,14}. The mechanism of far-field attraction to a wall may be in part explained by hydrodynamics \cite{15}. When a swimming cell is a so-called pusher, i.e.~its thrust is generated behind the cell body like spermatozoa or flagellated bacteria, the cell tends to direct towards the wall due to far-field hydrodynamic interactions. When a swimming cell is a puller, i.e.~when its thrust is generated in front of the cell body, on the other hand,  cells swimming parallel to the wall tend to be repelled by it hydrodynamically.

Some swimming microorganisms are seen to be trapped by  planar walls. Swimming \emph{E.~coli} cells are drawn to a stable circular trajectory, and the physical mechanism underlying the entrapment has been explained by hydrodynamic and steric effects \cite{16,17,18,19}. Spermatozoa can also be trapped by a planar wall, and the entrapment mechanism is again explained by hydrodynamic and steric effects \cite{20,21,22}. Such entrapment phenomena may reduce the spreading of cells near surfaces considerably. 

Other types of surface-related behaviour have also been reported for both biological and artificial swimmers.
{
	\emph{Chlamydomonas} algae cells scatter from a flat wall due to contact between its flagella and the surface and exhibit billard-like motion in polygonal geometries \cite{Spagnolie201733}. The collective behaviour of \emph{Bacillus subtilis} swarms and their interactions has been shown to depend on the geometry of the enclosing cavities \cite{wioland2016ferromagnetic}.
}
Chemically propelled rods tend to move on a surface along large circles with stochastic changes in the sign of the orbit curvature \cite{24} and may be captured around solid spheres of various radii and materials \cite{26}. 
Colloidal  particles rolling along a planar wall displayed a transition to macroscopic directed motion for large  populations \cite{25}. Swimming \emph{E.~coli} cells were seen to be entrapped along convex walls
provided their  curvature was sufficiently low \cite{27}. Theoretically, an analysis of the mechanism of entrapment  showed that a swimmer approaching a spherical colloid can be captured hydrodynamically when the colloid is larger than a critical size \cite{28}.

In nature many microorganisms live in environments subject to complex geometrical constraints. Bacteria in soil, for instance, swim in heterogeneous  granular matter, and their habitat is influenced by physical parameters including granular size and shape \cite{30,31,32,33,34}. While locomotion of organisms in granular media with particle size smaller than the body length has been investigated widely \cite{35,36}, only a few studies have addressed the spreading of microswimmers in complex geometries \cite{denis}.   A study on the motion of Janus particles in a two-dimensional (2D) periodic pattern of ellipsoidal pillars arranged in a triangular lattice  showed that microswimmers can steer even perpendicularly to an applied force in the periodic pattern \cite{37}. Swimming  Janus particles and \emph{E.~coli} cells in a two-dimensional colloidal crystal showed that  artificial particles orbit individual colloids with occasional hops while the circular orbits bacteria along planar walls were rectified into long, straight runs  \cite{38}. 

Given the potential environmental complexity encountered by both natural and artificial  swimmers, a lot remains to be explored. In  particular,  the role played by both the type of swimmer and  its  environment in dictating how the swimmers spread on average is a fundamental quantity of interest.  To this end, we consider in this paper a model system of swimming in a periodic environment.  The swimmer is a spherical squirmer \cite{39_lighthill,40_blake},  one of the simplest models of low-Reynolds number swimmers that allows us to compute the hydrodynamics precisely. The swimmer is then located in an infinite three-dimensional body-centred cubic (BCC) lattice of passive, rigid spheres of the same radius and its trajectory is computed using Stokesian dynamics \cite{41}. Running a large number of simulations, we are able  to address three main questions in our study: (1) which qualitative types of trajectories arise in periodic lattices, (2) how these depend on the strength of the swimmer and the packing density of the lattice and (3) how to explain these results from  hydrodynamic arguments.

\section{Model and Simulations}

\begin{figure}[t!]
	\centering
\includegraphics[width=0.45\textwidth]{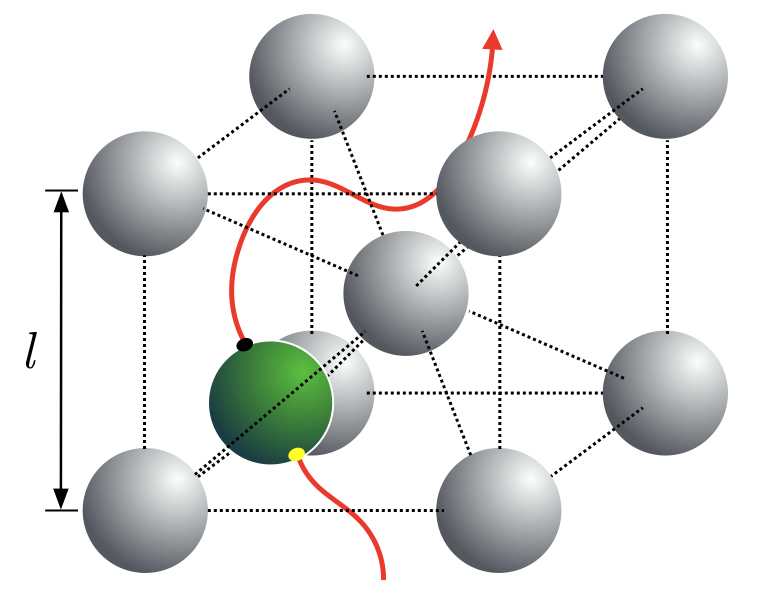}
\caption{Unit lattice cell: The squirmer (green sphere with front indicated in black and back in yellow) self-propels  between rigid spheres (grey) forming a BCC lattice of period  $l$.  The  computational domain consists of $4\times4\times2$ such unit lattice cells with periodic boundary conditions, containing a single squirmer.}
\label{Fig1}
\end{figure}

Our computations are carried using a modified version of the Stokesian Dynamics code used and validated in Ref.~\cite{41}. We consider a single spherical squirmer of radius 1 and 64 identical rigid spheres of the same radius 1 arranged as a rigid stationary BCC lattice. The flow satisfies the no-slip boundary condition on their surface. A unit lattice cell has side length $l$ and contains the equivalent to two lattice spheres, while the computational cell consists of $4\times 4\times 2$ unit cells, thus including the equivalent to 64 lattice spheres, plus the squirmer (see Fig.~\ref{Fig1}). At the boundary of the computational cell, periodic boundary conditions for the flow are applied. The lattice volume fraction, or packing density, $\phi$ is related to $l$ by $\phi=8\pi/3l^3$.

The squirmer induces an effective slip velocity tangential to its surface in order to propel forward. Specifically, in spherical coordinates measured from its centre, the velocity boundary conditions applied by the squirmer in the swimming frame are given by \cite{39_lighthill,40_blake}
\begin{equation}\label{squirmerBC}
u_r(1,\theta)=u_{\phi}(1,\theta)=0,\quad u_{\theta}(1,\theta)=\frac{3}{2}\sin\theta\left(1+\beta\cos\theta\right),
\end{equation}
where  $\theta$ is the azimuthal angle from the squirmer orientation $\mathbf{e}$ and $\beta$ is the so-called squirming parameter dictating the nature of the far-field flow. A swimmer with $\beta>0$ is a puller, like the alga {\it Chlamydomonas}, whereas one with $\beta<0$ is a pusher, similar to most flagellated bacteria and spermatozoa. 

In absence of the lattice, the boundary conditions in Eq.~\ref{squirmerBC} have been chosen such that an individual squirmer would swim with unit speed \cite{40_blake}. The characteristic time  used to non-dimensionalise the equations is thus the time scale it takes for one swimmer to swim is own radius. With this scaling, the  simulations uses a timestep of $10^{-3}$ and a total of $10^6$ steps, thus is equivalent to 1000 dimensionless time units. The viscosity of the fluid, $\eta$, is also scaled to unity.

In our simulations the squirmer is initialised at the centre of a face of a unit lattice cell, with either random or prescribed orientation. Note that through the periodic boundary conditions our model  corresponds to an infinite lattice with an infinite number of squirmers. However, the separation between them is large enough to make any hydrodynamic interactions negligible compared to those with the surrounding lattice spheres, giving an appropriate model for a single swimmer in an infinite lattice.

In what follows, we examine volume fractions between 1\% and 13\%. The theoretical maximum to allow space for a squirmer to pass through the lattice along one of its axes (i.e.~such that $l\geq4$) is $\phi=\pi/24\approx13.1\%$, so the examined parameter range includes the very dilute limit up to almost the densest possible lattices that the swimmer could theoretically traverse. We also vary the strength of the squirming parameter by considering the range $-3\leq\beta\leq2$ and thus include both pullers an pushers. 

\section{Computational results}

\subsection{Zoology of trajectories}

We ran a large number of simulations in our  parameter range $-3\leq\beta\leq2,\,1\%\leq\phi\leq13\%$ with random initial orientations. We  observe qualitatively  different features of the resulting trajectories originating from a complex interplay of both far-field and near-field interactions with the lattice (see illustration in Fig.~\ref{traj} and details below).  In order to  classify  the various different kinds of behaviour rigorously and systematically, we first introduce some  definitions. 

\subsubsection{Definition \#1: Turns vs.~straight}\label{sec:defstraight}

A squirmer is said to make a turn every time its orientation changes by more than $60^\circ$ from when it last made a turn (or from the initial orientation in case of the first turn). {
If a squirmer makes no turn during the entire simulation, or more than 500 time units elapse between two subsequent turns, then we say the trajectory is straight. The latter condition is added to discount transients occurring at the beginning of the simulation and to account for trajectories with very long straight segments, and the results are not sensitive to the exact value of the threshold, which has merely been chosen to be much larger than the time scale of scattering from the lattice.}
The angular threshold can be motivated by the typical behaviour observed in our numerical simulations: the swimmers do not perform U-turns, but often tend to wiggle around an otherwise straight path. The particular angular value we have chosen reflects these observations and has proven to capture the intuitive idea of a turn (as opposed to a wiggle) correctly in all  cases considered.  Note that the translational velocity direction and the orientation of the squirmer need not coincide and we may also have chosen this as  a criterion for the definition (although it did perform not as effectively as the  angular criterion). Note furthermore that our criterion does not distinguish between a sharp turn and a long curved trajectory; however the latter does not occur in our simulations and the swimmer only changes its orientation substantially in the neighbourhood of a lattice sphere.  

\subsubsection{Definition \#2: Random walks}

When swimmers are able to spread in the lattice without being trapped,  but do  not display straight trajectories, their behaviour is akin to that of a random walk with periods of straight swimming followed by reorientations after collisions with lattice spheres. When a swimmer follows a random walk, we measure the total  number of turns undergone by the swimmer over the entire course of the simulation (see below).

\subsubsection{Definition \#3: Trapped}

Writing the position of the centre of the squirmer as $\mathbf{r}(t)$, we define a squirmer to be trapped if there is a time $t^{\star}$ such that $|\mathbf{r}(t^{\star}+t)-\mathbf{r}(t^{\star})|\leq 5$ for all times $0<t<100$. This means that a swimmer is considered trapped if it does not move further than 5 radii in 20 times the time it would require to traverse that distance at full speed. The spatial threshold here is chosen to identify squirmers trapped by (and so remaining in contact with) a single lattice sphere, in which case the displacement is limited to slightly more than 4 radii. The temporal threshold is chosen to be much larger than the time required for the free case, and is not sensitive to the precise value as squirmers tend to get caught early in the simulations.

\subsubsection{Definition \#4: Stuck vs.~Orbits}
If a swimmer remains in  the proximity of a single lattice sphere, we introduce two further definitions to distinguish swimmers that  stay in place  from those that move on confined periodic or chaotic trajectories.  A trapped trajectory is called stuck if it remains straight, and orbiting if it does not.

\subsection{Phase diagram}

\begin{figure}[t!]
	\centering
\includegraphics[width=0.8\textwidth]{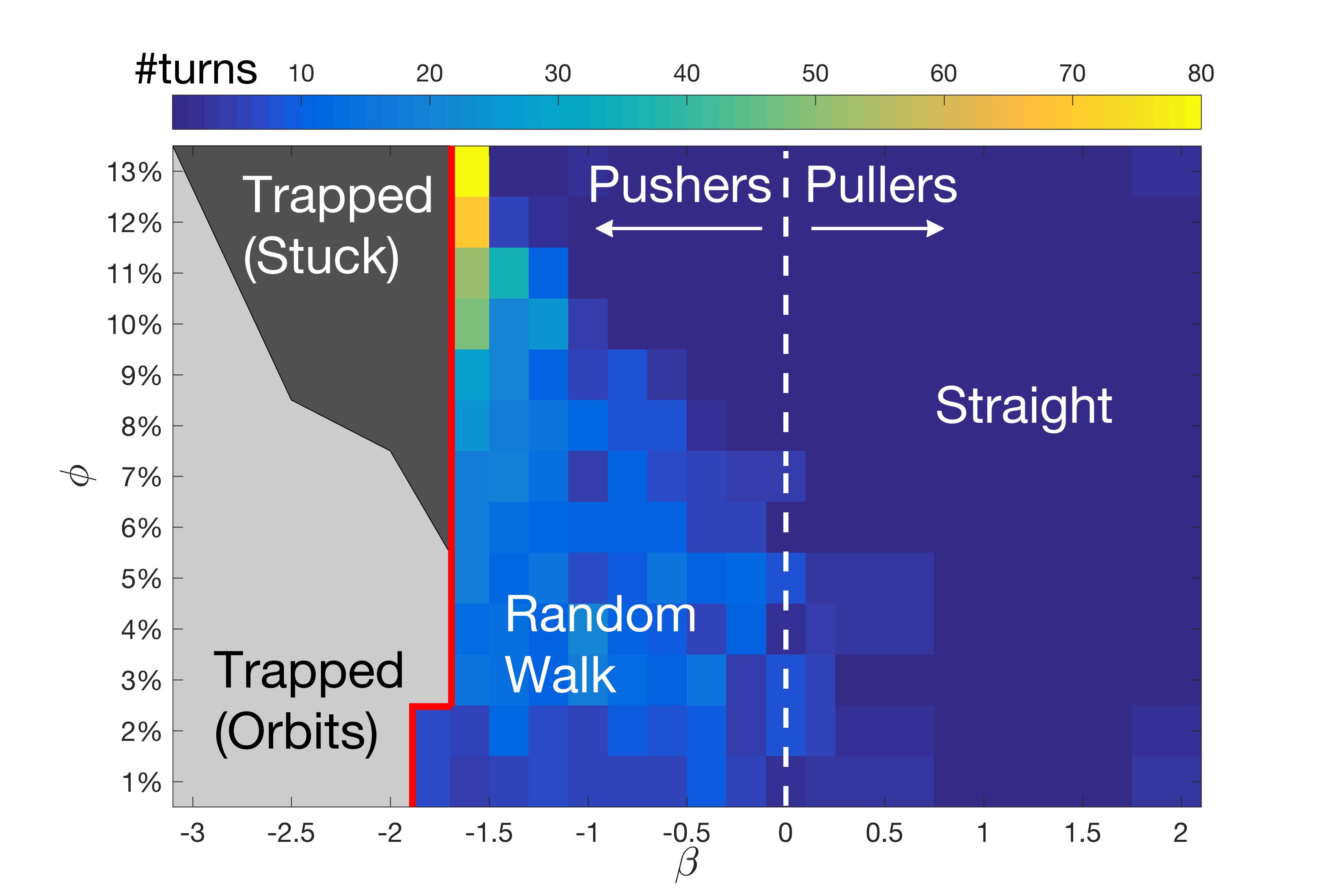}
\caption{Phase diagram in the $\beta-\phi$ plane diagram of active swimmer behaviour, based on test runs with random initial orientations, showing four qualitatively different kinds of trajectories. The red line indicates the transition between a  trapped behaviour for strong pushers to a random walk and straight trajectories for weak pushers and pullers. Coloured cells quantify the number of turns exhibited by a trajectory.}\label{phasediag}
\end{figure}

With these definitions, we can now summarise our computational results in a phase diagram. This is shown in Fig.~\ref{phasediag}  in the $\beta-\phi$ plane for squirmers with random initial orientations.  Four  different behaviours can be distinguished: straight trajectories; random walks (with a colour scheme capturing the number of turns occurring in total during the full computation); orbiting swimmers; and stuck swimmers. The red line around $\beta \approx -1.8 $ to $ -1.6$ provide the boundary between trapped swimmers (strong pushers)  and those which are not trapped (pullers and weak pushers).

\begin{figure}[t!]
	\centering
	\includegraphics[width=0.9\textwidth]{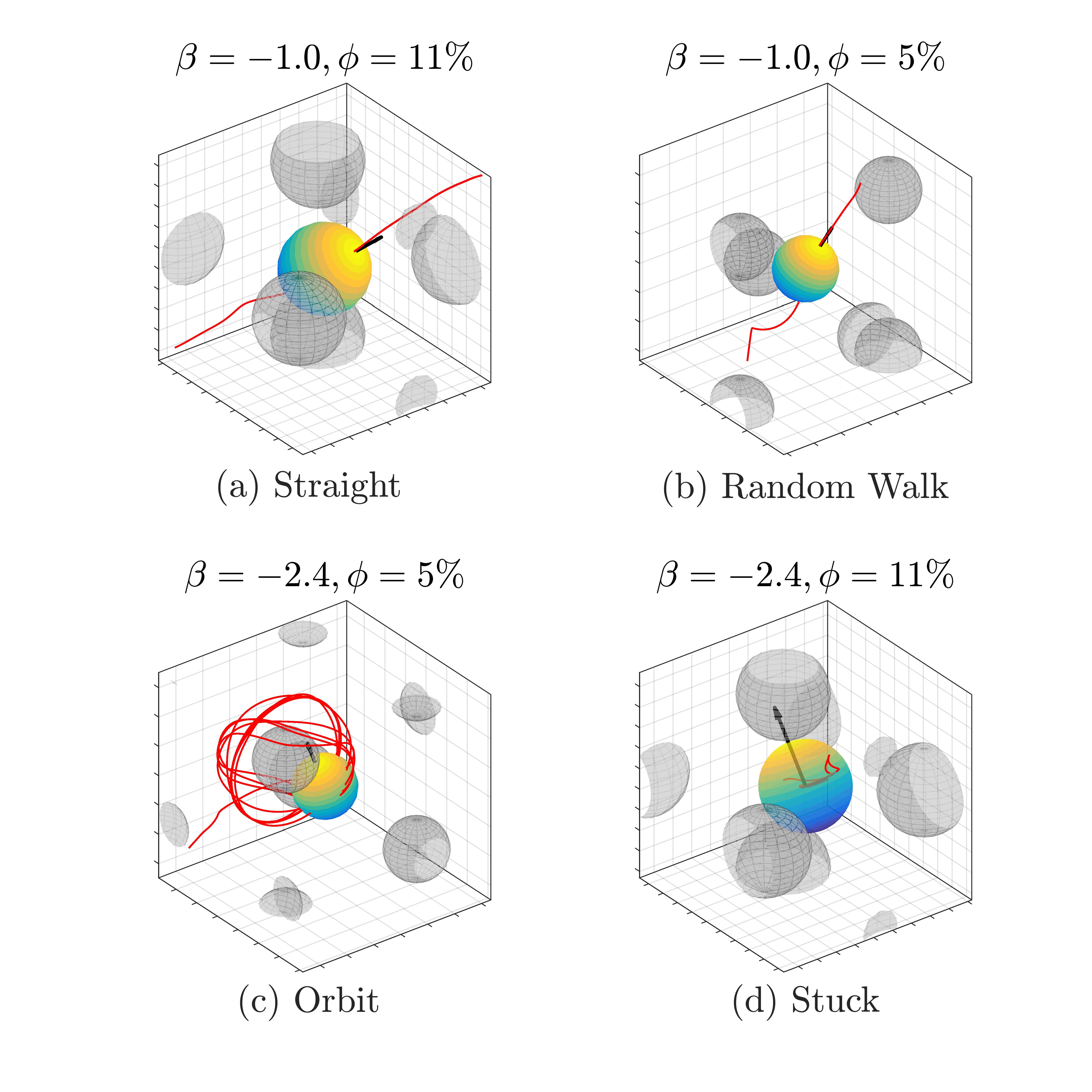}
	\caption{Sample trajectories for a swimmer moving in a BCC lattice for various values of the squirming parameter ($\beta$) and the lattice volume fraction ($\phi$):  
		(a) Weak pusher ($\beta=-1.0$) undergoing straight swimming ($\phi=11\%$); 
		(b) Weak pusher ($\beta=-1.0$) undergoing a random walk in the lattive ($\phi=5\%$); 
		(c) Strong pusher ($\beta=-2.4$) orbiting around of the spheres in the lattice ($\phi=5\%$); 
		(d) Strong pusher ($\beta=-2.4$)  stuck in the lattice ($\phi=11\%$).
	}\label{traj}
\end{figure}

\begin{figure}[t!]
	\centering
\begin{subfigure}[t]{0.4\textwidth}
		\centering
		\includegraphics[width=\textwidth]{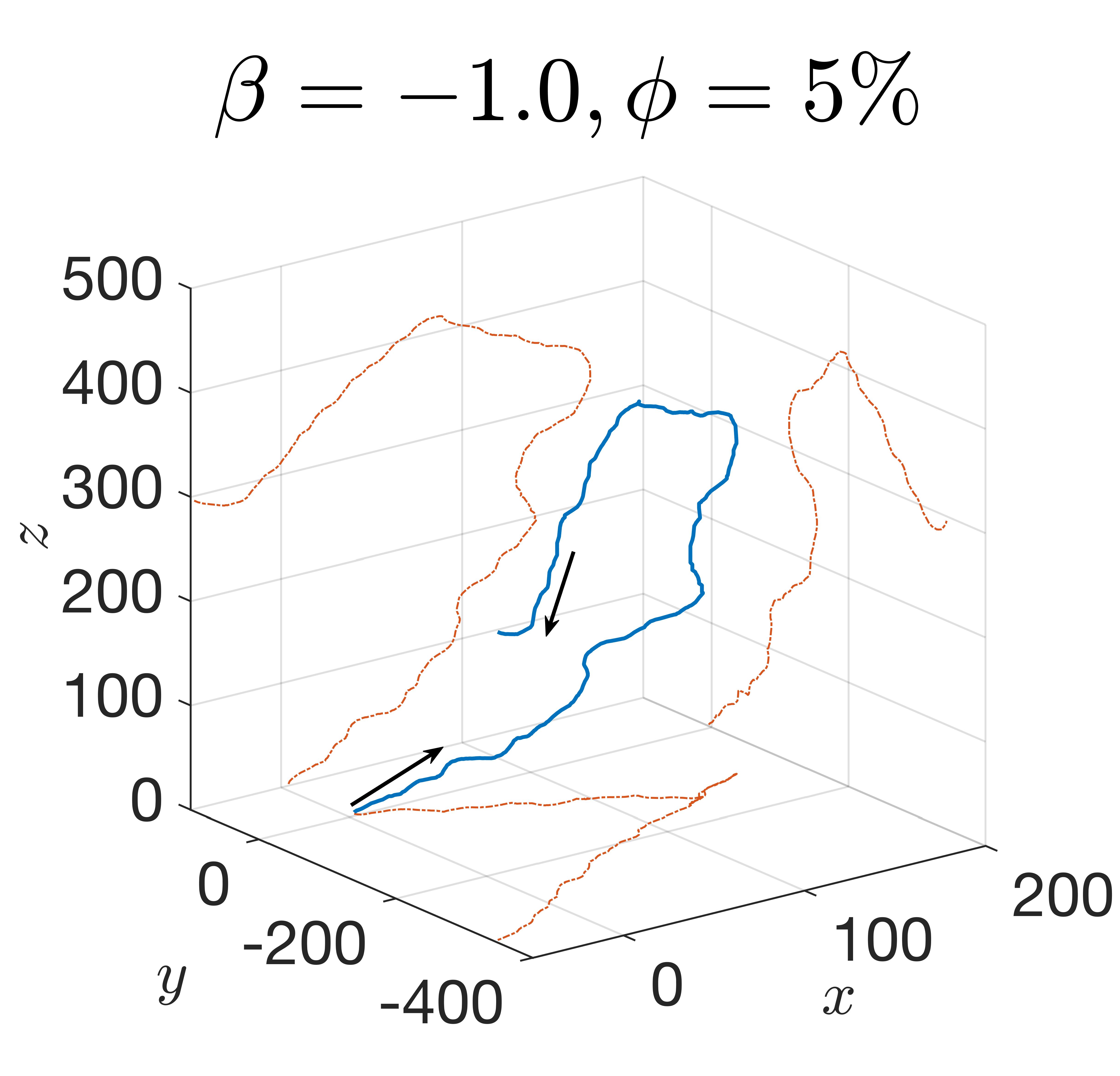}
		\caption{Random Walk}
	\end{subfigure}
~
	\begin{subfigure}[t]{0.4\textwidth}
		\centering
		\includegraphics[width=\textwidth]{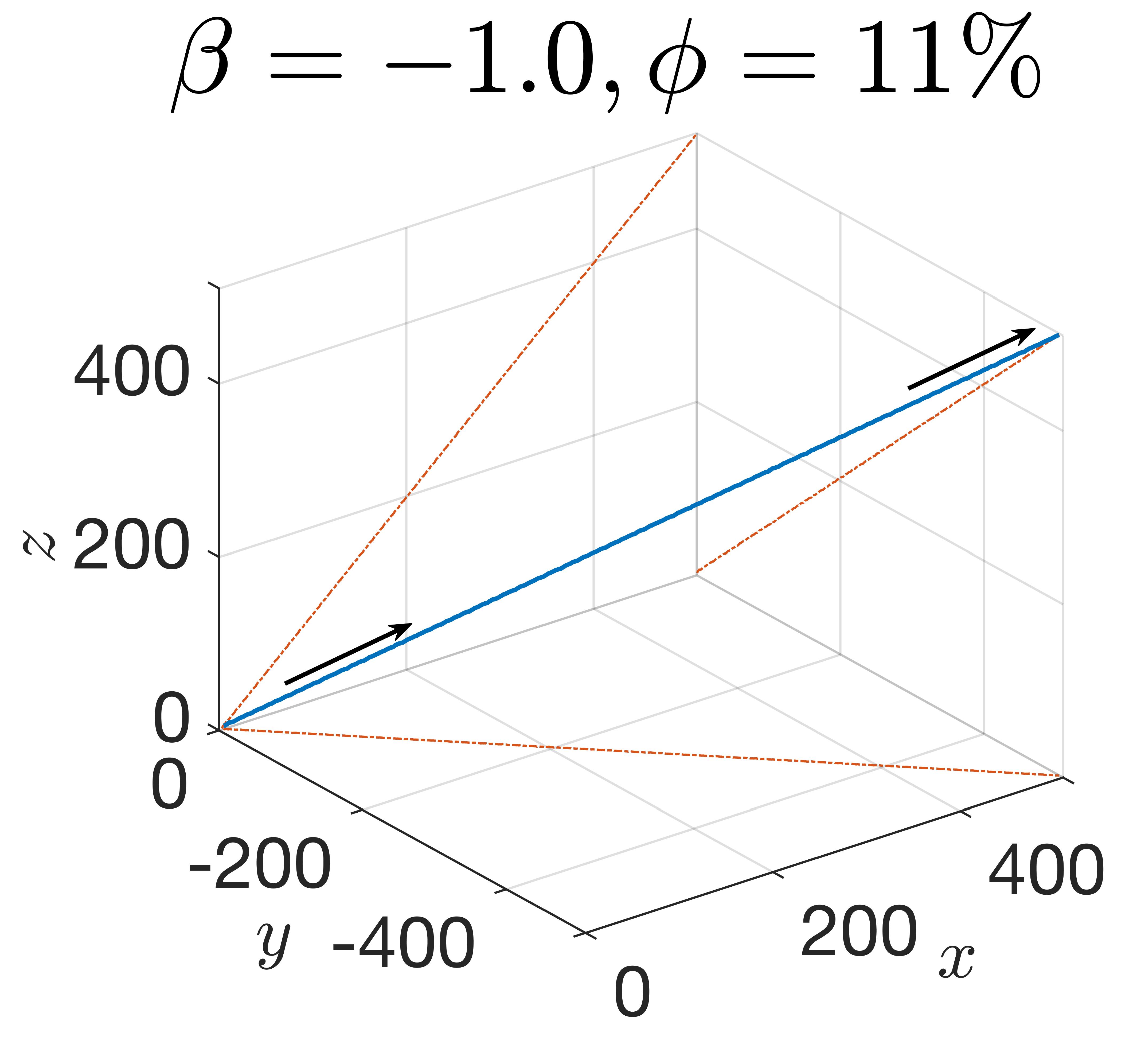}
		\caption{Straight}
	\end{subfigure}
	\caption{Long-time trajectories (blue solid  lines) of intermediate-strength pushers ($\beta=-1$) transitioning from random walks at low volume fraction (a, $\phi=5\%$) to straight swimming at high volume fractions (b, $\phi=11\%$). Shadows on the $x-y$, $x-z$ and $y-z$ planes are drawn to help the eye (red  dash-dotted lines), and arrows indicate the direction of motion (length scales are scaled by the squirmer radius).}
	\label{back}
\end{figure}

\subsubsection{Pullers}

All pullers considered in our simulations ($\beta>0$) move through the lattice in  nearly straight paths at almost unit speed. These paths are not limited to perfectly straight lines parallel to the lattice axes but includes also diagonal paths with small wiggles around obstacles. A simple far-field theoretical approach in \S\ref{sec:theory} will allow us to rationalise these observations. 
 
 \subsubsection{Pushers}
 
 For weak pushers ($-1.6\lesssim \beta<0$), similar straight trajectories are observed at small volume fractions, albeit with larger wiggles (Fig.~\ref{traj}a). However,  when the volume fraction becomes larger the active particles are deflected through interactions with the lattice, leading to  random walks (Fig.~\ref{traj}b).  Most of  these random walks are  very weak, with a maximum of  $\approx20$ turns over the course of the simulations, corresponding to a turn only every couple of lattice cells (see coloured cells in Fig.~\ref{phasediag}). Further we observe that the number of turns in the random walks increase with the volume fraction of the lattice ($\phi$) and with the strength of the flow imposed by the swimmer ($\beta$), as might be expected from intuition. In contrast, for very small values of $\phi$ the interactions become weaker and the number of turns reduces to a small non-zero value for a wide range of  $\beta$ values. Interestingly, for intermediate-strength pushers, the random trajectories return to straight swimming for very high values of $\phi$, as illustrated in Fig.~\ref{back}. 
 
When the pusher becomes stronger than a critical value, found to be approximately $\beta\lesssim-1.6$ in our simulations (and closer to $\beta\lesssim-1.8$ in the  very dilute lattice limit)  we observe a sharp transition to a trapped state in a manner which appears to be largely independent of the volume fraction in the lattice. In the trapped region, at small volume fractions, a swimmer orbits a single  lattice sphere (Fig.~\ref{traj}c) but becomes stuck for larger values of $\phi$ (Fig.~\ref{traj}d). The range of values of $\beta$ leading to a stuck swimmer  shrinks as $\beta$ is decreased further until all trajectories are orbits at $\beta\leq -3$. Interestingly, these orbits are nearly periodic circles for small volume fractions, but become  chaotic for larger values of  $\phi$, apparently through constraints by the surrounding lattice environment. In all cases,  orbiting squirmers are not oriented  tangentially to the lattice sphere, but always tend to be at a small acute angle with it that increases with $|\beta |$, effectively `sliding' around their captivators.

\subsubsection{Dependence on initial orientation}
\begin{figure}[t!]
	\centering
\includegraphics[width=0.6\textwidth]{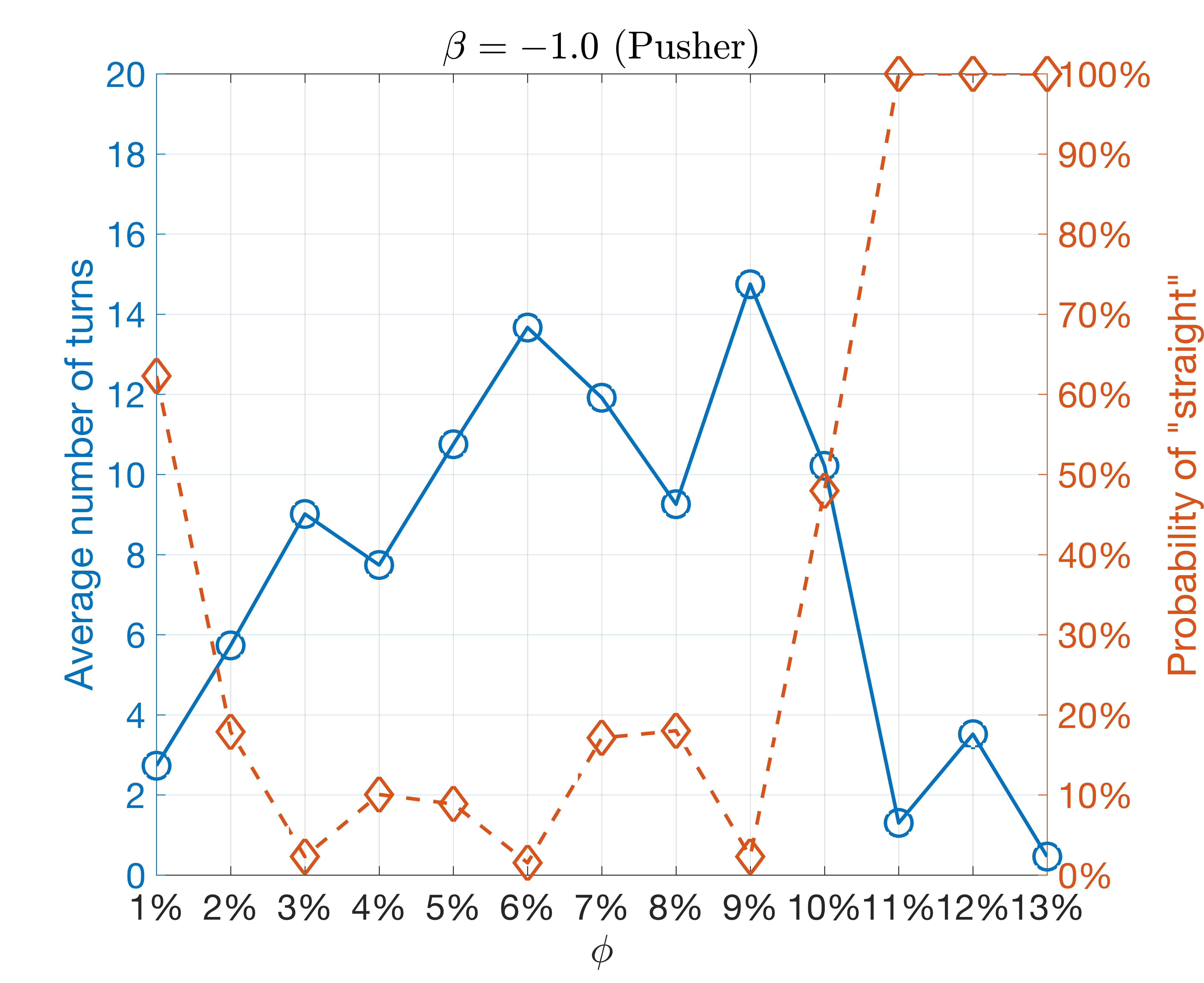}
\caption{Characteristics of trajectories for a weak pusher ($\beta=-1.0$) calculated using Lebedev quadrature from 12 sample initial orientations:  expected number of turns (blue solid line) and probability that a trajectory is straight (dashed red line). The transition from random walk to straight occurs at a volume fraction around $\phi=10\%$.}\label{beta-1}
\end{figure}

In order to analyse the dependence of the random walks on the volume fraction of the lattice  we performed multiple runs for selected values of $\beta$ and $\phi$ using different initial orientations of the swimmer. The space of orientations is suitable for Lebedev quadrature  so that expected values can be calculated for a randomly initialised squirmer \cite{lebedev1999quadrature}. We chose 17th-degree quadrature which ordinarily requires 110 sample orientations but this can be reduced to 12 by exploiting the  symmetries of the lattice, reducing computational effort substantially. Recognising that the trajectory depends very sensitively on the initial orientation, we emphasise that the analysis of these results should be interpreted heuristically.

In Fig.~\ref{beta-1} we plot for a given weak pusher ($\beta = -1.0$) the expected number of turns (blue solid line) and the probability that a trajectory is straight (dashed red line).  Strikingly, the probability to obtain a straight trajectory rises sharply to unity at volume fractions greater or equal to $10\%$, along with a corresponding sharp decrease in the number turns (that need not be zero, see \S\ref{sec:defstraight}). Likewise, we see that trajectories are on average more straight at very low volume fractions, which is what one would expect for the case of free swimmer.  Similar computations for values of $\beta$ in the trapped regime (strong pusher, not shown) also reveal that the average distance travelled by swimmers  through the lattice decreases as either the volume fraction and or the squirming parameter increase.

\section{Theory}
\label{sec:theory}

Our simulations reveal a number of remarkable features of the locomotion of an active particle in a lattice. In this section, we provide theoretical arguments in order to rationalise  these observations. We focus below on three main questions arising from the  phase diagram in Fig.~\ref{phasediag}: (1) Why are pullers able to swim straight while weak pushers tend to undergo random walks? (2) Why is there a threshold for pushers to be trapped? (3) How can we understand the transition, in the trapped regime, between an orbiting and a stuck state as a function of the volume fraction of the lattice?

\subsection{Straight swimming vs.~random walks}
 In order to understand why  pullers swim straight though the lattice  while weak pushers tend to undergo random walks, we use a far-field analysis of the hydrodynamic interactions between the swimming spheres and the stationary spheres in the lattice. In this limit, this can be accomplished using the method of reflections \cite{BookHappelBrenner,kim2005microhydrodynamics}. Below we show analytically that  squirmers swimming through a channel of spheres are stabilised and destabilised along their straight trajectory in approximately equal intervals but with varying strength and in a manner that changes sign with the squirming parameter. 
  
The leading-order flow induced by the squirmer is that of a stresslet (force-dipole) which can be written as \cite{Batchelor1970,blake1971spherical} 
\begin{equation} \label{eq:dipole}
\mathbf{u}(\mathbf{r}) = -\frac{3}{4}\beta\left(3\frac{\mathbf{r}(\mathbf{e}\cdot\mathbf{r})^2}{r^5}-\frac{\mathbf{r}}{r^3}\right),
\end{equation}
where the squirmer is instantaneously centred at the origin and has orientation $\mathbf{e}$.  The leading-order hydrodynamic  response to this flow by a lattice sphere located at $\mathbf{r_0}$ is a Stokeslet (point force) that can be calculated using Fax\`en relations \cite{kim2005microhydrodynamics}. The response of the squirmer to this Stokeslet is then a perturbation, $\mathbf{u}_{\rm Stokes}$,  to its swimming  velocity as
\begin{equation}
\mathbf{u}_{\rm Stokes}=\frac{9}{8}\beta(3\cos^2\theta-1)\frac{\mathbf{r}_0}{r_0^4}+\mathcal{O}\left(r_0^{-4}\right),\quad \cos\theta:=\mathbf{e}\cdot\mathbf{\hat{r}}_0,
\end{equation}
with  higher-order terms due to further flow singularities and their responses.  Importantly, the  sign of this velocity perturbation depends both on the squirming parameter, $\beta$, and on the angle that $\mathbf{e}$ makes with the displacement vector to the lattice sphere; this is a consequence of the dipolar nature of the flow in Eq.~\ref{eq:dipole}. From this, it is clear that the geometry of the lattice and the position and orientation of the squirmer within it determine its behaviour in a very non-trivial fashion.

\begin{figure}[t!]
	\centering
	\includegraphics[width=0.7\textwidth]{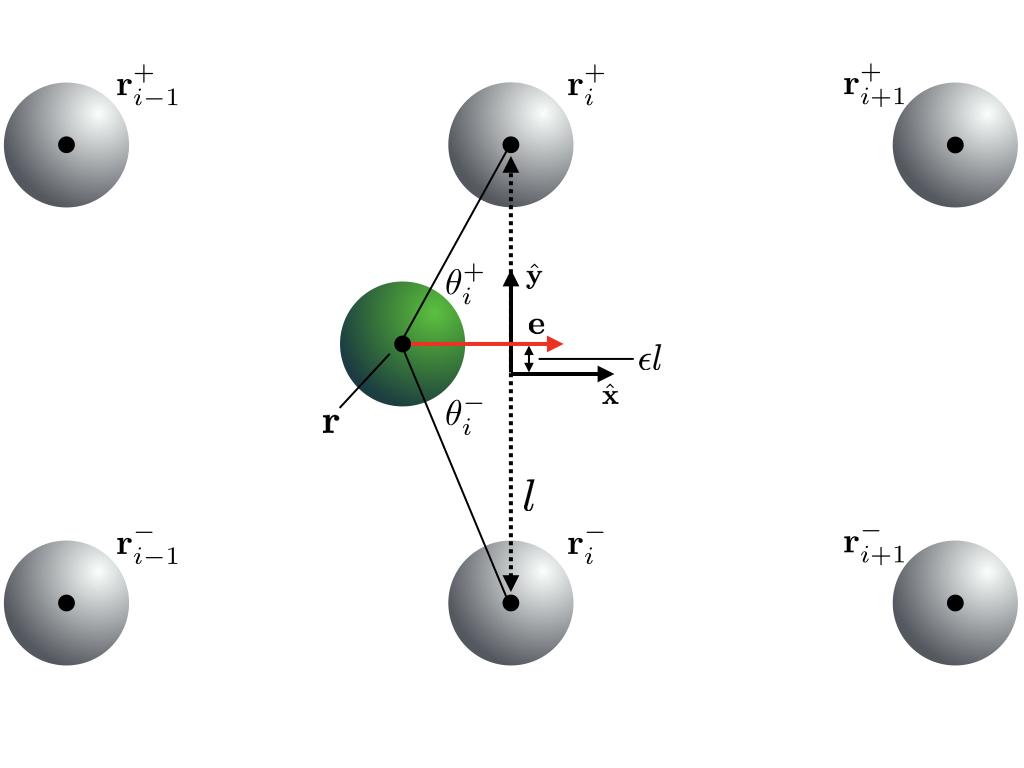}
	\caption{Simplified model used to explain the transition from straight swimming (pullers) to random walks (pushers). The squirmer moves through a planar channel of width $l$ made of unit rigid spheres. The position of the swimmer is denoted $\mathbf{r}=(xl,\epsilon l)$ and those of the fixed spheres by $\mathbf{r}_i^\pm=il\hat{\mathbf{x}}\pm l/2\hat{\mathbf{y}}$ where $i$ is an integer. $\theta_i^\pm$ denotes the angle between $\mathbf{e}$, $\mathbf{r}$ and $\mathbf{r}_i^\pm$.}\label{setup22}
\end{figure}

Further progress can be made however by considering a simplified scenario, illustrated in Fig.~\ref{setup22}.  Consider a squirmer confined to a two-dimensional plane moving through a `channel' of lattice spheres positioned at $\mathbf{r}_i^\pm=il\hat{\mathbf{x}}\pm l/2\hat{\mathbf{y}}$ where $i$ is an integer. Such a channel is a common feature of regular lattices such as cubic, BCC and FCC, which can all be thought of as assembled from interlaced channels at different orientations.
{In particular, by linearity of the flow in the dilute limit the results obtained from this model will give qualitative predictions for the three-dimensional case at sufficiently low volume fractions.}

Since the Stokeslet responses are directed towards the centre of the squirmer, they do not exert any net torque on it. For this reason the rate of change of $\mathbf{e}$ is with $\mathcal{O}(l^{-5})$ at least two orders of magnitude smaller than the rate of change of $\epsilon$. For the purpose of the argument below, we may therefore assume that the orientation of a swimmer, $\mathbf{e}$, remains unchanged. We focus on the case $\mathbf{e}=\hat{\mathbf{x}}$ and quantify the role of hydrodynamic interactions on the motion of the sphere. The squirmer position is denoted by $\mathbf{r}=(x(t)l,\epsilon(t) l)$ where $|\epsilon|\ll 1$. Neglecting all but the leading-order interactions we obtain the dynamical system for  $(x,\epsilon)$ as
\begin{equation}\label{perteq}
\frac{{\rm d}x}{{\rm d}\tau}=1+\mathcal{O}(\epsilon), \quad\frac{{\rm d}\epsilon}{{\rm d}\tau}\approx\frac{9\beta}{8l^3}\sum_{i,\pm} \frac{3\cos^2\theta_i^\pm-1}{|\mathbf{r}_i^\pm-\mathbf{r}|^3}\sin\theta_i^\pm,
\end{equation}
where $\tau=t/l$ and $\theta_i^\pm$ is the angle between the vectors $\mathbf{e}$ and $\mathbf{r}_i^\pm-\mathbf{r}$. Upon linearising in $\epsilon$ the terms in the sum we obtain the differential equation for the off-axis location of the swimmer as
\begin{equation}\label{eq:DS}
\frac{{\rm d}\epsilon}{{\rm d}x}=-\frac{36\beta}{l^3}\left\{\sum_{n=-\infty}^{\infty}\frac{3-52(n-x)^2+32(n-x)^4}{[1+4(n-x)^2]^4}\right\}\epsilon+\mathcal{O}(\epsilon^2).
\end{equation}
which integrates to
\begin{equation}
	\ln{\left(\frac{\epsilon}{\epsilon_0}\right)} \approx -\frac{36\beta}{l^3}\sum_{n=-\infty}^{\infty}\left\{\frac{11(x-n)+16(x-n)^5}{4(1+4(x-n)^2)^3}+\frac{1}{8}\tan^{-1}\left(2(x-n)\right)\right\}
\end{equation}
Each term in the sum represents a kink in $\ln\epsilon$ centred around $x=n$ of size $\pi/8$ and width $\mathcal{O}(1)$. Taking the average over unit intervals we can therefore deduce that
\begin{equation}
\left\langle \ln \left(\epsilon / \epsilon_0\right) \right\rangle \approx -\frac{9\pi\beta}{2l^3}x=-\frac{27}{16}\beta\phi \tau \quad \Rightarrow \quad \epsilon(\tau) \sim e^{-\left(27/16\right)\beta\phi \tau}.\label{logav}
\end{equation}
From this we see that puller trajectories ($\beta>0$) are stable, whereas pushers ($\beta<0$) are not.

\begin{figure}[t!]
	\centering
	\includegraphics[width=0.7\textwidth]{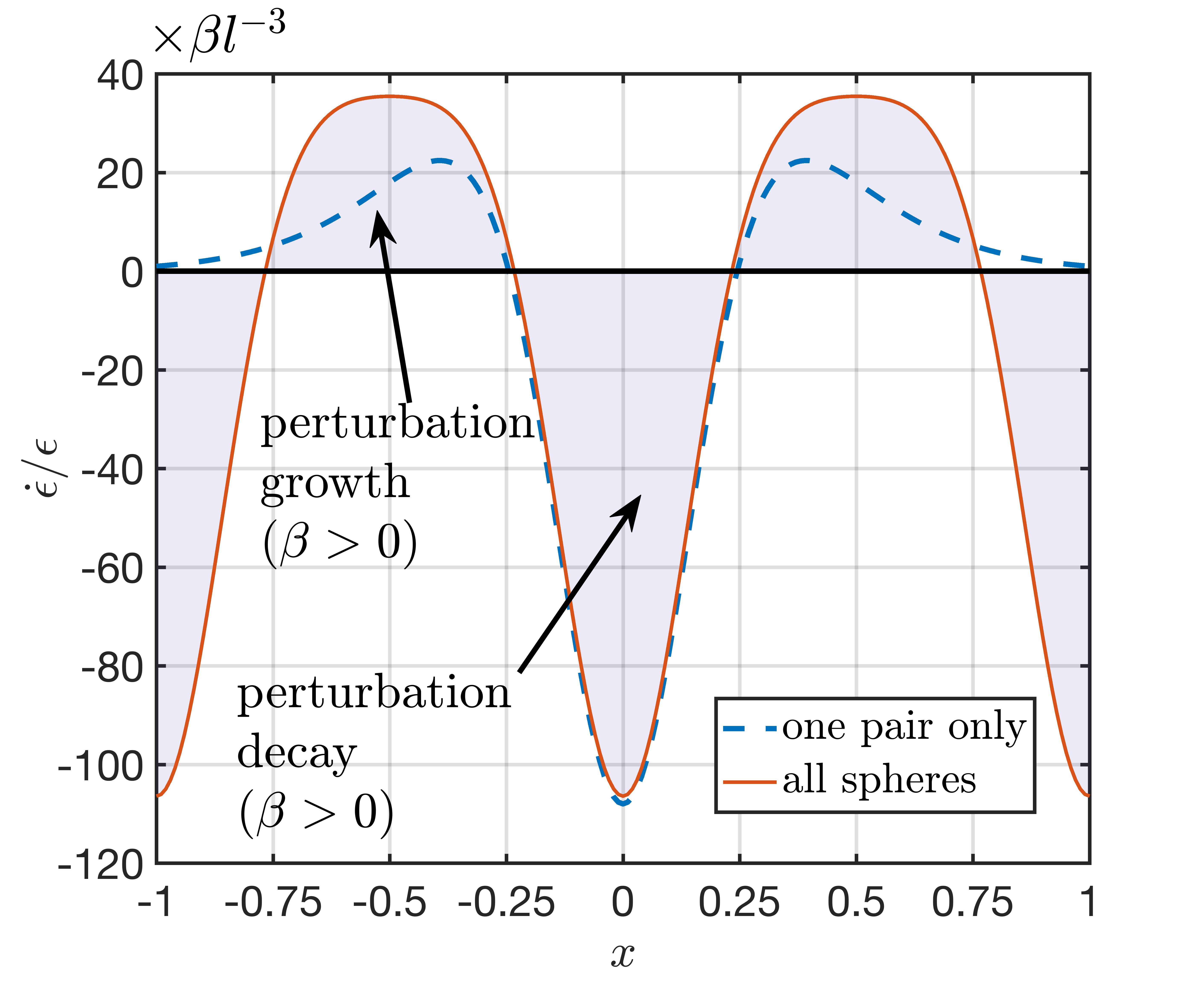}
	\caption{Instantaneous growth/decay of a perturbation $\epsilon$ away from a trajectory through the middle of a two-dimensional channel of rigid spheres. Results are shown in the  case of a puller ($\beta>0$); for a pusher the stability properties are reversed via a change of sign. The contribution from a single $\pm$-pair of spheres is displayed for illustration purposes (blue dashed line) while the total effect of all spheres  is a superposition of such pair effects (red solid  line).}\label{setup2}
\end{figure}

\begin{figure}[t!]
	\centering
	\includegraphics[width=0.7\textwidth]{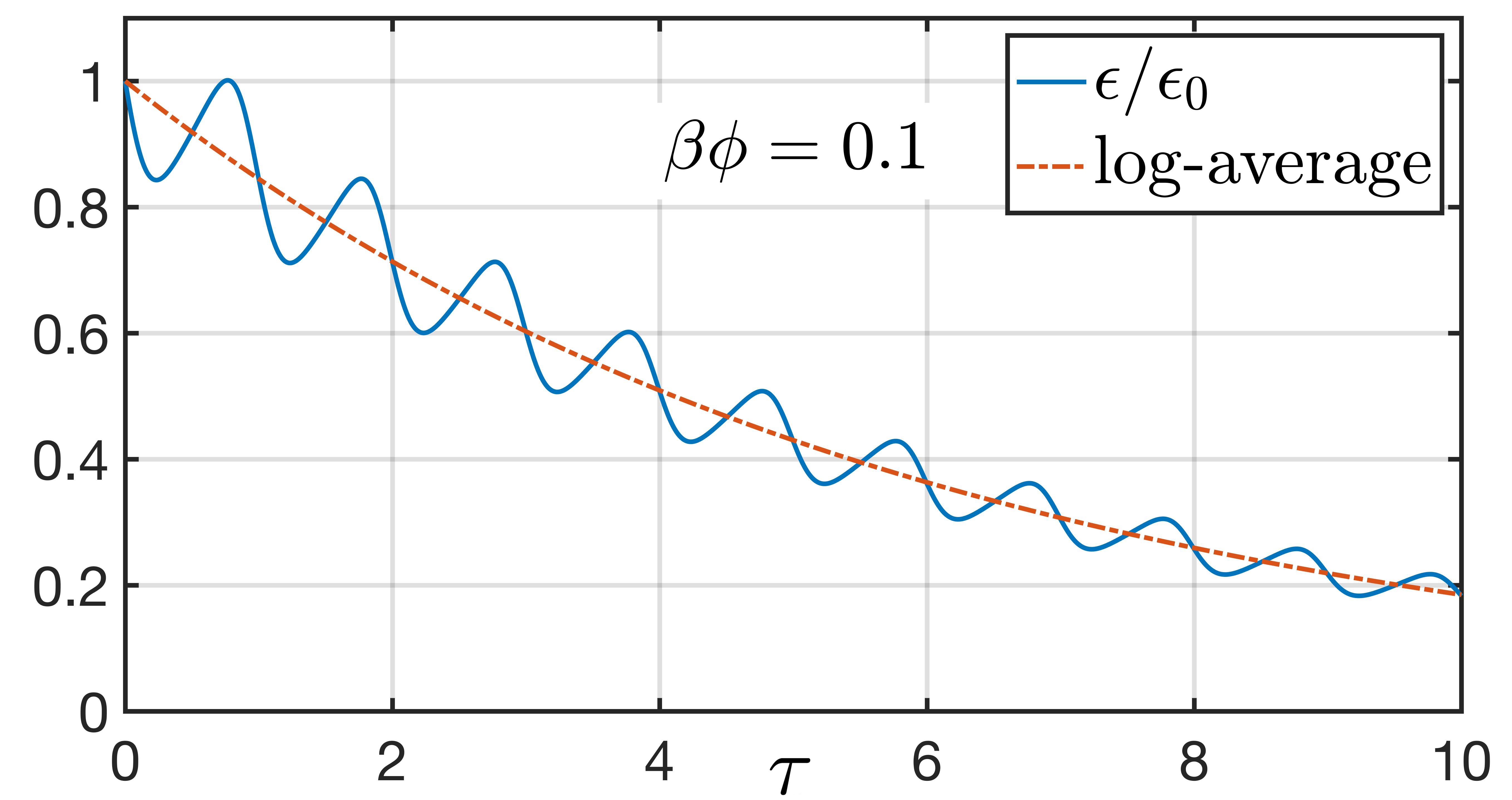}
	\caption{Decay of a perturbation for the off-axis position of the swimmer in a two-dimensional channel of rigid spheres (setup in Fig.~\ref{setup22}) in the case of puller ($\beta>0$) and $\beta\phi=0.1$ over 10 unit lattice lengths (blue solid line). The shape of the curve is generic for a puller. While the deviation of the swimmer away from the axis undergoes periods of both growth and decay, overall the latter dominates the dynamics, leading to exponential decay of the perturbation on average (Eq.~\ref{logav}, red dashed line). }\label{perturbation}
\end{figure}

In Fig.~\ref{setup2} we plot the effect of a single pair of spheres at $y=\pm l/2$, corresponding to one term in the series ($n=0$), on a puller squirmer $(\beta >0)$ depending on its position relative to them (dashed blue line). We also plot the influence of all spheres in the channel (solid red line). We note that any contributions from spheres more than $\approx0.75l$ away from the swimmer are vanishingly small compared to the closer ones, a result self-consistent with our neglecting of any sphere above and below the channel, which thus allows us to generalise our heuristic argument to the full lattice. As the swimmer moves through the lattice, the perturbation of its position away from the central axis of the channel, $\epsilon$, undergoes alternating periods of growth (positive regions in Fig.~\ref{setup2}) and decay (negative regions).  We observe that the perturbations grow in one direction when the swimmer is roughly between a $\pm$-pair of spheres and grow in the other when it is roughly in the middle between two such pairs. Importantly, the rate of change is considerably greater in the decay region than in the growth region. This is illustrated in Fig.~\ref{perturbation} where we plot the spatial variation of the perturbation, $\epsilon$,  as the swimmer makes its way through the lattice, as well as the log-average from Eq.\ \ref{logav}. As a consequence, the perturbation undergoes an oscillating decay for a puller and the straight trajectory is therefore stable, consistent with our numerical simulations. In contrast, for a pusher the decay/growth regions in Fig.~\ref{setup2} are reversed and the swimmer is ultimately unstable, as observed in our simulations at low volume fractions.
Furthermore we note that the force with which the squirmer is stabilised or destabilised  in this dilute model scales as $\sim \beta\phi$. Therefore the effect becomes very small at low volume fractions; in particular,  when the 
 orientation of
 the squirmer is not aligned with an axis of the lattice, trajectories become more complex and are impacted by  scattering from lattice spheres.

Conversely, as $\phi$ is increased for pushers, there is less space for them to turn and reorient, forcing them on an effectively straight trajectory. This is what is observed in Fig.~\ref{traj}a. The clear jump in the probability that pushers go straight in Fig.~\ref{beta-1} can therefore be explained by the existence of a critical volume fraction $\phi_c$ increasing with $|\beta|$,  above which the squirmers cannot reorient quickly enough to leave an effectively straight path through the lattice.

\subsection{Unbounded vs.~bounded trajectories}
Similar far-field considerations may be used to obtain an estimate for the threshold in pusher strength beyond which all swimmers are  trapped. For this, we make use of a recent paper which analysed the swimming of force dipoles near rigid spheres using the method of hydrodynamic images  \cite{28}. In that study, it was shown that  for a pusher  swimmer of unit radius  creating a dipolar flow of strength $\alpha$, trapping would occur along colloids whose radius, $A$, exceed a critical value  $A_c$ given by
\begin{equation}
A_c = \frac{64}{9\alpha^2}\cdot
\end{equation}
Swimmers near colloids for which $A<A_c$ will continue swimming (possibly with some scattering) while in the case $A>A_c$  pushers would all be predicted to be trapped.

Since the passive spheres of the lattice are of radius $1$, this means that trapping would be predicted to occur for strong pushers, $|\alpha | > |\alpha_c|$, with 
\begin{equation}
\alpha_c = \left(\frac{64}{9}\right)^{1/2} = \frac{8}{3}\cdot
\end{equation}
In our model, Eq.~\ref{eq:dipole} indicates that the strength of the dipolar flow is given by 
\begin{equation}
\alpha = -\frac{3}{4}\beta,
\end{equation}
and thus trapping is predicted to occur for this far-field model for strong pushers satisfying
\begin{equation}\label{ffbetac}
\beta < \beta_c,\quad \beta_c=-\frac{32}{9}\approx-3.56.
\end{equation}
This number is off by about a factor of two from the results of our numerical simulations ($\beta_{c,\rm num}\approx -1.8$). Since the theoretical prediction in Ref.~\cite{28} was derived in the case of a unique swimmer-sphere pair in the far field, and thus neglects hydrodynamic interactions with other spheres and all near-field effects, it is striking  that the prediction remains  robust and may also be used to predict the swimming behaviour in a lattice.

\subsection{Stuck swimmer vs.~orbiting}
So far we provided theoretical arguments for two of the main numerical observations, namely the qualitative difference between pushers and pullers as well as the transition, for the case of pushers, between a trapped and non-trapped regime. We now address the difference, within the trapped region, between swimmers which effectively remain stuck and those which continuously orbit around of the lattice spheres. In this case, the dynamics can only be explain through an interplay between generic far field hydrodynamic interactions with the lattice (as outlined above) and near-field hydrodynamic forces which necessarily  depend on the details of the squirmer actuation.

In order to capture near-field effects, we now derive an argument based on lubrication theory (i.e.~the equations of hydrodynamics valid in thin geometries~\cite{Batchelor_book})  to derive an expression for the torque experienced by a squirmer near an inert sphere as a function of  its orientation (assuming that we are beyond the threshold for bounded dynamics). The distinct stuck vs.~orbiting behaviour would then arise depending on whether lubrication (high volume fraction) or far-field effects (low volume fraction) dominate.
 
\begin{figure}[t!]
\centering
\includegraphics[width=0.4\textwidth]{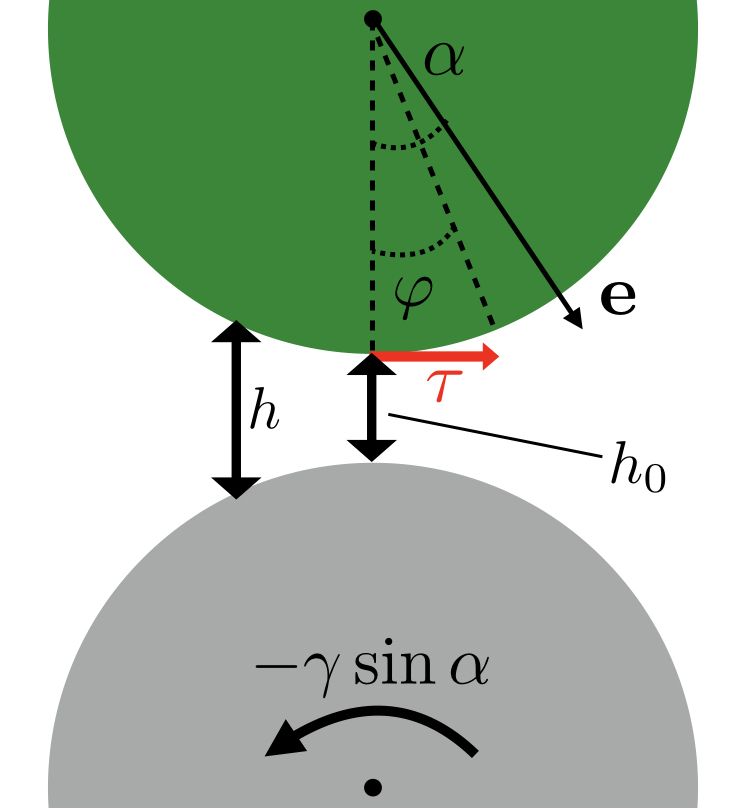}
\caption{Sketch of the two-dimensional lubrication argument giving rise to a viscous torque, $\tau$, acting on the squirmer (green, top) sliding around an inert lattice `sphere' (grey, bottom), as viewed in the frame where the squirmer is stationary and the sphere rotates with an angular velocity of $-\gamma\sin\alpha$ where $\alpha$ is the angle that the squirmer orientation $\mathbf{e}$ makes to the normal to the sphere and $\gamma$ a constant. For an angle $\varphi$ from the normal $h(\varphi)$ denotes the gap width, which has its minimum $h_0$ at $\phi=0$.}\label{lubr}
\end{figure}

To simplify calculations, we consider the two-dimensional case of two cylinders (`spheres'), which although they might exhibit different scalings for the forces and torques, would capture the same flow physics (see sketch in Fig.~\ref{lubr}). Because the dynamics are Stokesian there is no inertia, and hence a squirmer sliding around a sphere at constant velocity is completely equivalent to having a stationary squirmer next to a uniformly rotating sphere. We work in this frame in order to directly apply the squirming boundary condition, 
Eq.~\ref{squirmerBC}. We next assume that the squirmer remains at a constant distance $h_0$ from the inert `sphere' surface and that its swimming direction, $\mathbf{e}$, makes an acute angle $\alpha$ with the normal to it. 
The gap width is thus written as
\begin{equation}
h(\varphi)=h_0+2(1-\cos\varphi),
\end{equation}
where $\varphi$ is the angle between the normal and a point on the squirmer.
 We assume that the speed at which the surface is rotating is constant and write it as $u_{\rm slide}=-\gamma\sin\alpha$ where $\gamma$ is a constant between 0 and 1 that reflects the fact that it is slowed down by friction compared to the unit speed of an unobstructed swimmer. 
 
 From Eq.~\ref{squirmerBC} we have  that the tangential velocity on the squirmer is given by
 \begin{equation}
u_{\rm Squirmer}=-3/2\sin(\alpha-\varphi)(1+\beta\cos(\alpha-\varphi))
\end{equation}
in the direction of increasing $\varphi$ (see Fig.~\ref{lubr}). In the lubrication limit the flow in the gap is a pure shear flow and therefore the total  tangential stress on the squirmer is given by
\begin{eqnarray}
\tau & =& -\int_{-\varphi_0}^{\varphi_0}\frac{u_{{\rm Squirmer}}-u_{{\rm slide}}}{h}\cos^2\varphi\,{\rm d}\varphi \\
&=&-\int_{-\varphi_0}^{\varphi_0}\left(\frac{-\frac{3}{2}\sin(\alpha-\varphi)(1+\beta\cos(\alpha-\varphi))-\left(-\gamma\sin\alpha\right)}{h}\right)\cos^2\varphi\,{\rm d}\varphi ,
\end{eqnarray}
where $\varphi_0$ is a cut-off angle for the region in which lubrication theory is valid ($\sqrt{h}\ll1$) and the factor of $\cos^2\varphi$ arises from the need to project integrand and integration measure into the direction perpendicular to the gap. Carrying out the integration leads to
\begin{equation}\label{torque}
\tau = \sin\alpha \frac{3\pi}{2\sqrt{h_0}}\left(1-\frac{2}{3}\gamma+\beta\cos\alpha \right),
\end{equation}
at leading order in $h_0$ as $h_0\to0$. We note that $\tau\sim h_0^{-1/2}$ in this two-dimensional setting while for spheres one finds $\tau\sim\ln h_0^{-1}$ \cite{GOLDMAN1967637,ishikawa2006hydrodynamic}.

{
This traction results in a torque on the squirmer and we are interested in the consequences of this torque on the dynamics of trapped swimmers. Analysing Eq.~ \ref{torque} for equilibrium points in $\alpha$ we find that $\alpha=0$ is unstable for $\beta>\beta_c$ and stable for $\beta<\beta_c$ where
\begin{equation}\label{nfbetac}
	\beta_c=-1+\frac{2}{3}\gamma.
\end{equation}
and $0<\gamma<1$ implies $-1<\beta_c<-1/3$. Compared with Eq.~ \ref{ffbetac} we see that the threshold for entrapment predicted from considering near field effects only is much lower than that expected from far field effects. As the lubrication forces scale rather weakly with $h_0$ in three dimensions and higher order effects from the global geometry will matter, we conclude that a theoretical prediction for the critical squirmer strength should lie between those two extremes, as is the case with our numerical observed value of $\beta_c\approx-1.8$.
}

We  propose that  the sliding behaviour can be understood as the consequence of a balance between far- and near-field hydrodynamic effects, shifting $\alpha$ to a non-zero equilibrium value that increases with the squirmer strength (i.e.~with $|\beta|$). For  strong pushers, the far-field hydrodynamic interactions dominates, explaining why there are more orbiting and fewer stuck trajectories. Additionally, at high volume fractions interactions with other lattice spheres start to become important. The net effect of these appears to be repulsion once a squirmer has been trapped. When the strength of the swimmer is not too strong and the volume fraction of the lattice is sufficiently large, the repulsion is strong enough to reduce the equilibrium value of $\alpha$ to a very small value, resulting in the behaviour classified as `stuck'. In contrast, when the squirming strength is sufficiently large, the swimmer gets deflected from its near-circular trajectory around its captivator, giving rise to the observed  `orbiting' behaviour.

\section{Conclusion}

Using extensive numerical simulations we computed the trajectories of spherical squirmers through BCC lattices for a  broad range of both the squirming parameter, $\beta$, and the lattice packing density, $\phi$. We  found a rich array of different qualitative behaviours, ranging from straight trajectories over weakly diffusive random walks to instant entrapment by the lattice. 

We further demonstrated that much of this behaviour can be explained, at least heuristically, using a combination of relatively simple hydrodynamic near-field and far-field arguments for the squirmer and its interactions with individual  spheres in the  lattice. Specifically, pullers tend to be stabilised and pushers destabilised on trajectories far from lattice elements. Close to lattice spheres, near-field and far-field hydrodynamic torques compete. There is a clear threshold which is almost independent of the lattice volume fraction and a strong pusher above it is trapped, whereas a weak pusher below it is deflected but ultimately released back into the lattice. This may then result either in a trajectory resembling a weak random walk or, if too constrained by the lattice packing density, again a straight path. For a given squirming strength, there appears to be a clear threshold in packing density above which all trajectories are straight regardless of initial conditions.

In the parameter range for which the swimmer performs a random walk, it would be desirable to quantify the effective diffusive motion of the active particles. However, because reorientation events  occur very infrequently, it is currently too computationally expensive to quantify it accurately from our simulations. A recent study has investigated this for a random Lorentz gas \cite{denis}, and future work with more powerful numerical algorithms may shed more light on this issue taking hydrodynamics into account.  
{{Similarly, the effects of lattice geometry warrant further investigation: while test runs in a cubic lattice reveal the same generic behaviour of squirmers in the far field, the question to what extent the stuck--orbiting and trapped--free thresholds depend on the lattice geometry remains to be answered.}}

Future studies could address how  additional properties of the squirmer can affect the dynamics, for example potential  phoretic effects. In a setup with a chemically-patterned lattice, a passive particle may propel through phoretic interactions, or in the case of a weakly charged lattice a charged swimmer may be affected through electromagnetic effects. Practical applications in which our results may be put into use include the design of porous filters for swimmers of different types and strengths.

\section*{Acknowledgements}
This work was supported by a Rouse Ball Studentship by Trinity College to AC, a JSPS Grant-in-Aid for Scientific Research (A) (17H00853) to TI and an ERC Grant PhyMeBa (682754) to EL.

\section*{References}
\bibliographystyle{unsrt}
\bibliography{bibliography}

\end{document}